\documentclass[12pt]{report}
\usepackage{epsfig}
\usepackage{units}
\usepackage{amsmath,amssymb}
\usepackage{multirow}
\usepackage{rotating}
\usepackage{graphicx}
\usepackage{algorithm}
\usepackage{algorithmic}

\begin{document}
\thispagestyle{empty}
\begin{center}
\textbf{\large SCHOOL OF ELECTRICAL ENGINEERING
AND TELECOMMUNICATION\\ University of New South Wales, Australia}\\[2cm]
{\addtolength{\baselineskip}{0.5cm}
\textbf{\Huge
Gaussian Processes Techniques for Wireless Communications} \\[0.5cm]
}
{\Large by}\\[0.5cm]
\textit{\huge
Chong Han} \\[1.5cm]
{\Large
Thesis submitted as a requirement for the degree\\
Bachelor of Engineering (Telecommunications)\\[2ex]
\vfill
Submitted: Oct $12^{th}, 2010$\hfill

Supervisor: Prof. Jinhong Yuan\hfill

\vspace*{-1cm}
}
\end{center}

\begin{abstract}
Bayesian filtering is a general framework for recursively estimating the state of a dynamical system. Classical solutions such that Kalman filter and Particle filter are introduced in this report. Gaussian processes have been introduced as a non-parametric technique for system estimation from supervision learning. For the thesis project, we intend to propose a new, general methodology for inference and learning in non-linear state-space models probabilistically incorporating with the Gaussian process model estimation.
\end{abstract}

\tableofcontents

\chapter{Introduction}\label{ch:intro}
\begin{flushleft}
This report summarizes the work achieved in the first half of the entire thesis. Essentially, it includes the studying of a classic system model which has been widely used in many fields and its possible solutions within different scenarios. This work can be regarded as a preparation step. In the next half of the project, we will address a practical problem with the aid of these methodologies.
\newline
\newline
In overview,
chapter~\ref{ch:Bayesian} introduces the classic state-space model and its generic solution Bayesian approach.
Nevertheless, due to the integrals intractability in practice, chapter~\ref{ch:KF} describes Kalman filter for the linear state-space model while chapter~\ref{ch:PF} reveals the Particle filter methods for the more realistic non-linear model.
\newline
All of these methods rely on the condition that the state-space model information is deterministic but in many cases, we deal with the situation with uncertain model structure.
Rather than deciding the model relates to a specific model, chapter~\ref{ch:GP} includes the concept of a Gaussian process, the Gaussian process regression approach and supervision learning of the hyperparameters. In the end, materials that have been referred to are included in Bibliography and in the Appendix, MatLab codes for Kalman filter, Particle filter and Gaussian process regression are provided.
\end{flushleft}

\chapter{Bayesian Approach}\label{ch:Bayesian}
\section{Introduction - State-space Model}
We consider probabilistic state-space models of the form
\begin{align}
x_k &= f_k(x_{k-1},u_{k-1},w_{k-1}){\label{SSM1}}\\
z_k &= h_k(x_{k},u_{k},v_{k}){\label{SSM2}}
\end{align}
where
\begin{itemize}
\item $f$: state transition or evolution function
\item $x_k,x_{k-1}$: current and previous \textit{state}
\item $u_{k-1}$: known input
\item $w_{k-1}$: state noise
\item $h$: measurement function
\item $z_k$: observation
\item $u_{k}$: known input
\item $v_{k}$: measurement noise
\end{itemize}
Our aim is to provide a sequence of optimal (with respect to the minimum mean square error criterion (\textit{MMSE}) estimates $\hat{x_k}$ of a process. The true state is hidden and the information available upon which our estimate rely is a set of measurements (or observations) $\{z_{0:k}\}$. \\
The state-space model is used in the fields of channel estimation in wireless communications. For example the Autoregressive (AR) of first order is a well-accepted approximation of the Jake's channel update model~\cite{wang2002verifying}. Moreover, state-space model has been also widely used to predict economic data in finance, track positions in control system and recover image or speech in signal processing.
\section{Bayesian Approach}
With the fact that the evolution of the state follows a Markov Process of order one (Equation~\ref{SSM1}), a Bayesian approach solves the filtering problem in a sequential manner by incorporating all observations into account. This amounts to calulating the posterior distribution of the state $p(x_k|z_{0:k})$ at each instant $k$. Assume that we have the access to the known previous state $p(x_{k-1}|z_{0:k-1})$. The idea of forming the required posterior of the next state is to combine the previous state information with $p(x_k|x_{k-1})$ from the state transition and $p(z_k|x_k)$. This prediction step is processed before the new observation coming. So as $z_k$ is obtained, we advance to the next step to update our prior estimate. In overall, the recursion proceeds in two stages, prediction and update as shown following.
\subsection{Bayesian Approach - Prediction}
The a prior estimate of the posterior distribution at $k$ is given by
\begin{align*}
p(x_{k}|z_{1:k-1}) &= \int p(x_k,x_{k-1}|z_{1:k-1})dx_{k-1}\\
&= \int p(x_k|x_{k-1},z_{0:k-1})p(x_{k-1}|z_{1:k-1})dx_{k-1}\\
&= \int p(x_k|x_{k-1})p(x_{k-1}|z_{1:k-1})dx_{k-1}
\end{align*}
where we used the Markov property $p(x_k|x_{k-1},z_{0:k-1})=p(x_k|x_{k-1})$ and the prediction result is known as the Chapman-Kolmogorov equation~\cite{arulampalam2002tutorial}.
\subsection{Bayesian Approach - Update}
By incorporating the new observation with the a prior estimate, we can update the posterior distribution as
\begin{align*}
p(x_k|z_{0:k}) &= p(x_k|z_k,z_{0:k-1})\\
&= \frac{p(x_k,z_k,z_{0:k-1})}{p(z_k,z_{0:k-1})}\\
&= \frac{p(x_k,z_k|z_{0:k-1})}{p(z_k|z_{0:k-1})}\\
&= \frac{p(z_k|x_k,z_{0:k-1})p(x_k|z_{0:k-1})}{p(z_k|z_{0:k-1})}\\
&= \frac{p(z_k|x_k)p(x_{k}|z_{1:k-1})}{p(z_k|z_{1:k-1})}
\end{align*}
where $p(z_k|z_{1:k-1})=\int p(z_{k}|x_{k})p(x_{k}|z_{1:k-1})dx_{k}$ is the normalizing constant(evidence or marginal likelihood).
\newline
\section{Summary}
In theory, this Bayesian approach utilizes all the information available and it can provide a closed form solution to the problem.
However in practice, intractable integrals and awkward equations may often occur and they are impossible to be evaluated analytically~\cite{doucet2009tutorial}.

\chapter{Kalman Filter}\label{ch:KF}
\section{Introduction}
Kalman filter is an algorithm that produces a MMSE estimator of the state process recursively. It requires the assumptions such that~\cite{arulampalam2002tutorial}
\begin{itemize}
\item Noises $w$ and $v$ are i.i.d. drawn from Gaussian distribution with known parameters
\item Evolution function $f$ and update function $h$ are both linear
\end{itemize}
Thus, if the previous state $p(x_{k-1}|z_{0:k-1})$ is Gaussian, then at the next time step $p(x_{k}|z_{0:k})$ is Gaussian as well. So the state-space model equations~\ref{SSM1} and~\ref{SSM2} can be rewritten as
\begin{align*}
x_k &= F_{k}x_{k-1}+w_{k-1}\\
z_k &= H_{k}x_{k}+v_{k}
\end{align*}
where $F_k$ and $H_k$ are known matrices defining the linear functions. In addition, we define $w_{k-1}$ has zero mean and covariance $Q_{k-1}$; $v_k$ has zero mean and covariance $R_k$.
\section{Kalman Filter - Algorithm}
Suppose that we have been up to one state $k-1$ and we have the access to $p(x_{k-1}|z_{1:k-1})=\mathcal{N}(m_{k-1|k-1},P_{k-1|k-1})$, the recursive algorithm under the Bayesian framework
consists of two steps, prediction and update. In this section, we will briefly introduce the Kalman filter algorithm.
\subsection{Prediction}
Inserting the previous state into the evolution equation~\ref{SSM1}, we can find a prior distribution of the state $x_k$ as
\begin{equation}{\label{eq:KF prediction}}
p(x_k|z_{1:k-1}) = \mathcal{N}(m_{k|k-1},P_{k|k-1})
\end{equation}
where
\begin{align*}
m_{k|k-1} &= F_{k}m_{k-1|k-1}\\
P_{k|k-1} &= Q_{k-1}+F_{k}P_{k-1|k-1}F_k^T
\end{align*}
\subsection{Update}
As we obtain the new observations $z_k$, we are able to update the posterior distribution as follows.
\begin{equation}{\label{eq:KF update}}
p(x_k|z_{1:k}) = \mathcal{N}(m_{k|k},P_{k|k})
\end{equation}
where
\begin{align*}
m_{k|k} &= m_{k|k-1}+K_k(z_k-H_{k}m_{k|k-1}) and\\
P_{k|k} &= P_{k|k-1}-K_{k}H_{k}P_{k|k-1}
\end{align*}
and the Kalman gain is $$K_k = P_{k|k-1}H_k^T(H_{k}P_{k|k-1}H_k^T+R_k)^{-1}$$.
\section{Simulation}
Consider an $AR(2)$ example as follows.
\begin{align*}
x_k &= 2cos(2\pi f)x_{k-1}-x_{k-2}\\
z_k &= x_k+v_k
\end{align*}
where $v_k$ is a Gaussian noise with zero mean and variance $var\_v$.
This problem can be rewritten in state-space form such that
\begin{align*}
\begin{bmatrix}
x_k \\
x_{k-1}
\end{bmatrix}
&=
\begin{bmatrix}
2cos(2\pi f) & -1 \\
1 & 0
\end{bmatrix}
\begin{bmatrix}
x_{k-1} \\
x_{k-2}
\end{bmatrix}
\\
z_k &= \begin{bmatrix}
1 & 0
\end{bmatrix}
\begin{bmatrix}
x_k \\
x_{k-1}
\end{bmatrix}
+ v_k
\end{align*}
Using the Kalman filter algorithm, we obtain the simulation result in the following figure and the MatLab codes are included in the Appendix $1$ and $2$.
\begin{figure}[h!]{\label{fig:KF_AR2}}
\centering
\includegraphics[width=1\textwidth]{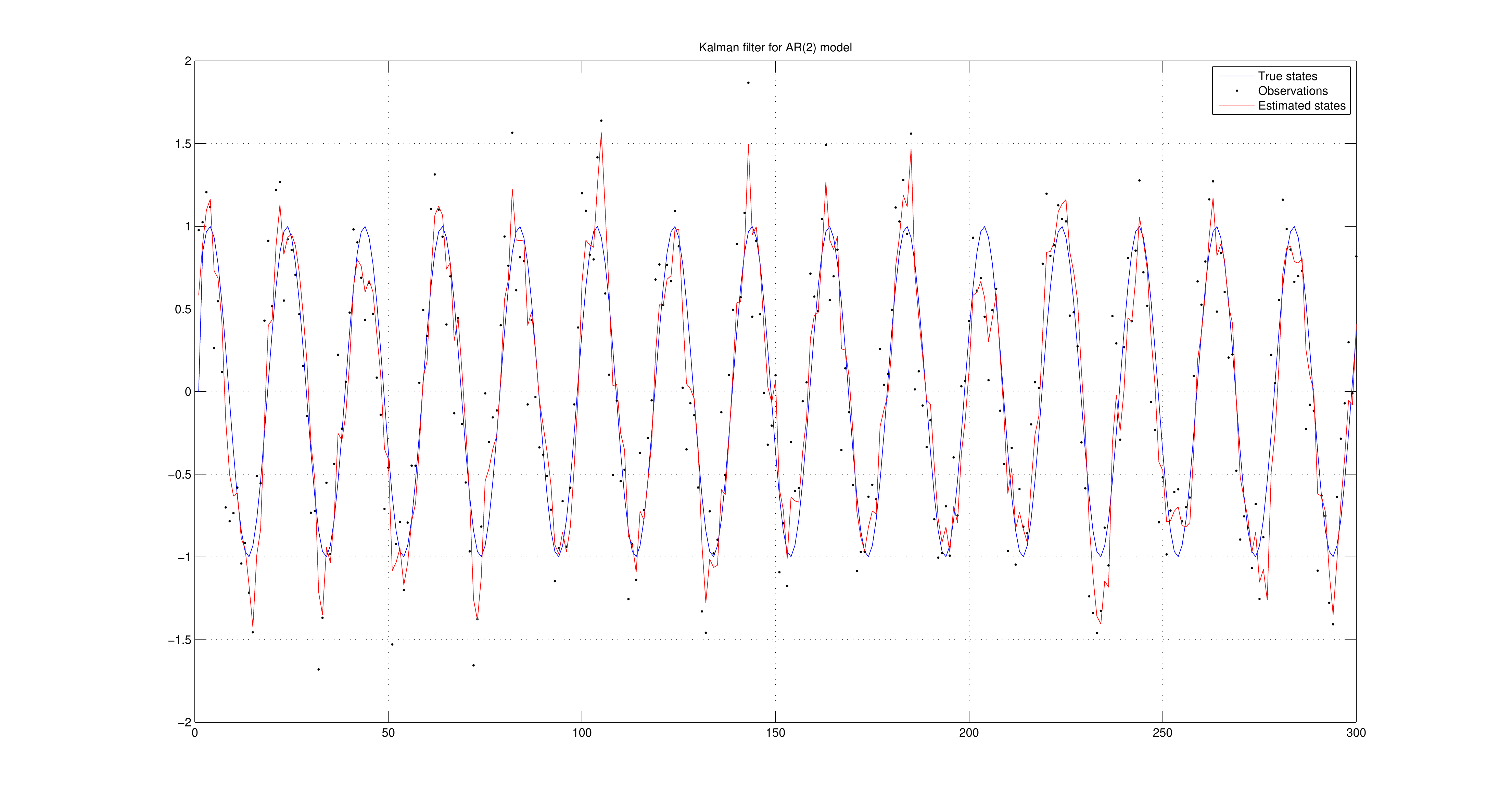}
\caption{Kalman filter for $AR2$ SSM}
\end{figure}

\section{Summary}
With the assumptions held, Kalman filter provides the optimal solution in this linear Gaussian environment.
However when the assumptions of system linearity and Gaussian noise are not available, Kalman filter does not perform well.
In the next chapter, we will describe an algorithm that performs superior for the non-linear state-space model problems.
\chapter{Particle Filter}\label{ch:PF}
\section{Introduction}
For linear Gaussian state-space model, Kalman filter is served as an optimal recursive estimator under the Bayesian framework. However, what if the state-space model is not restricted as linear and Gaussian? Instead of Kalman filter and its approximation~\cite{grewal2001kalman}, we will introduce particle filtering methods to solve these estimation problems numerically in an online manner - recursively as observations become available.
\newline
Particle filters perform sequential Monte Carlo (SMC) estimation based on point mass (or particles) representation of probability densities. Thus the key idea to resolve this state-space model probelm is to represent the state posterior density function by a set of random samples (also known as particles) with associated weights. As the number of samples approaches infinity, particle filter result approaches the optimal Bayesian solution.
\section{Monte Carlo Integration}
Monte Carlo integration is the basis of SMC methods. Suppose we want to numerically evaluate a multidimensional integral
$$I=\int g(x)dx$$
Monte Carlo (MC) methods for numerical integration can be factorized $g(x)=f(x)\pi(x)$ in such a way that $\pi(x)$ is interpreted as a probability density satisfying $\pi(x)\geq0$ and $\int \pi(x)dx=1$. Drawing $N\gg1$ samples ${x^i,i=1,\ldots,N}$ distributed according to $\pi(x)$, the MC estimate of integral
$$I=\int f(x)\pi(x)dx$$
with the sample mean
$$I_N=\frac{1}{N}\sum_{i=1}^N f(x^i)$$
If the samples $x^i$ are independent then $I_N$ is an unbiased estimate and according to the law of large numbers $I_N$ converges to the true value of $I$.
\newline
Ideally we want to generate samples directly from $\pi(x)$ but in the context of filtering, $\pi(x)$ is the posterior whose samples we cannot obtain. Instead, we perform the sampling from a density $q(x)$ named as the importance or proposal density. Following the principle of Importance Sampling, this proposal density is an approximation density to the true density $\pi(x)$. In this case, the integral of $I$ can be rearranged as $$I=\int f(x)\pi(x)dx=\int f(x)\frac{\pi(x)}{q(x)}q(x)dx$$
 A Monte Carlo estimate of $I$ is computed by generating independent samples distributed according to $q(x)$ and forming the weighted sum
$$I_N=\frac{1}{N}\sum_{i=1}^N f(x^i)\tilde{w}(x^i)$$
where $\tilde{w}(x^i)=\frac{\pi(x^i)}{q(x^i)}$ are the importance weights and they can be normalized
$$w(x^i)=\frac{\tilde{w}(x^i)}{\sum_{j=1}^N \tilde{w}(x^j)}$$
Then we estimate $I_N$ using the normalized importance weights to evaluate the integral
$$I_N=\frac{1}{N}\sum_{i=1}^N f(x^i)w(x^i)$$
\section{Sequential Importance Sampling}
The sequential importance sampling (SIS) is a Monte Carlo method upon which most sequential MC filters are relied on. This sequential Monte Carlo (SMC) approach is known variously as bootstrap filtering, the condensation algorithm, particle filtering, interacting particle approximations, and survival of the fittest.~\cite{arulampalam2002tutorial} Essentially it is a technique to implement a recursive Bayesian filter with the aid of Monte Carlo simulations. The principle is to represent the posterior density function by a summation of a set of random samples (particles) with associated weights and the tasks can be simplified to be finding the proper samples and their corresponding weights. By the law of large number, this approximation approaches to the real posterior density function as the number of samples becomes very large. In another word,the SIS filter becomes the optimal Bayesion estimator when $N_s$ approaches infinity.
\newline
Before developing the details of the algorithm, we introduce $\{x_{0:k}^i,w_k^i\}$ to be a random measure that characterizes the posterior pdf $p(x_{0:k}|z_{1:k})$ where $\{x_{0:k}^i, i=1,\ldots,N_s\}$ is a set of support points (particles) with associated weights $\{w_{k}^i, i=1,\ldots,N_s\}$. The weights are normalized such that $\sum_{i}w_k^i=1$. Then the posterior density at $k$ can be approximated as
$$p(x_k|z_{1:k})\approx \sum _{i=1}^{N_s}w_k^i\delta (x_k-x_k^i)$$
This is interpreted as the weighted approximation of the true posterior $p(x_{0:k}|z_{1:k})$. The normalized weights $w_k^i$ are chosen based on the principle of \textit{Importance Sampling}. Therefore, if the samples $x_{0:k}^i$ were drawn from an importance density $q(x_{0:k}|z_{1:k})$, then the weights become
$$w_k^i\propto \frac{p(x_k^i|z_{k})}{q(x_k^i|z_k)}$$
If the importance density can be factorized like this
$$q(x_{0:k}|z_{1:k})\triangleq q(x_k|x_{0:k-1},z_{1:k})q(x_{0:k-1}|z_{1:k-1})$$
then we can obtain samples $x_{0:k}^i \sim q(x_{0:k}|z_{1:k})$ by augmenting each of the existing samples $x_{0:k-1}^i \sim q(x_{0:k-1}|z_{1:k-1})$ with the new state $x_{k}^i \sim q(x_{k}|x_{0:k-1},z_{1:k})$. The full posterior distribution can be rearranged as
\begin{align*}
p(x_{0:k}|z_{1:k}) &=\frac{p(z_k|x_{0:k},z_{1:k-1})p(x_{0:k}|z_{1:k-1})}{p(z_{k}|z_{1:k-1})}\\
&=\frac{p(z_k|x_{0:k},z_{1:k-1})p(x_k|x_{0:k-1},z_{1:k-1})p(x_{0:k-1}|z_{1:k-1})}{p(z_{k}|z_{1:k-1})}\\
&=\frac{p(z_k|x_{k})p(x_{k}|x_{k-1})p(x_{0:k-1}|z_{1:k-1})}{p(z_{k}|z_{1:k-1})}\\
&\propto p(z_k|x_{k})p(x_{k}|x_{k-1})p(x_{0:k-1}|z_{1:k-1})
\end{align*}
and the weight update is
\begin{align*}
w_k^i &\propto \frac{p(z_k|x_k^i)p(x_k^i|x_{k-1}^i)p(x_{0:k-1}^i|z_{1:k-1})}{q(x_k^i|x_{0:k-1}^i,z_{1:k})q(x_{0:k-1}^i|z_{1:k-1})}\\
 &= w_{k-1}^i\frac{p(z_k|x_k^i)p(x_k^i|x_{k-1}^i)}{q(x_k^i|x_{0:k-1}^i,z_{1:k})}
\end{align*}
Furthermore, if $q(x_k|x_{0:k-1},z_{1:k})=q(x_k|x_{k-1},z_k)$ then the importance density appears only related to $x_{k-1}$
and $z_k$. This turns out to be useful when only a filtered estimate of $p(x_k|z_{1:k})$ (incomplete posterior) is required at each step.
In this case, the weight update becomes
$$w_k^i\propto w_{k-1}^i\frac{p(z_k|x_k^i)p(x_k^i|x_{k-1}^i)}{q(x_k^i|x_{k-1}^i,z_k)}$$
and finally, the prediction of posterior filtered density is approximated as
$$p(x_k|z_{1:k})\approx \sum _{i=1}^{N_s}w_k^i\delta (x_k-x_k^i)$$
In summary, the SIS algorithm is formed by recursive propagation of importance weights $w_k^i$ and particles $x_k^i$ as the sequential observation is obtained at each step. The algorithm is described in Algorithm~\ref{al:SIS}.~\cite{Ristic}
\begin{algorithm}
\caption{SIS Particle Filter}
\label{al:SIS}
\begin{algorithmic}
\STATE $[{\mathbf{x_k^i},w_k^i}]=SIS[{\mathbf{x_{k-1}}^i,w_{k-1}^i},\mathbf{z_k}]$
\begin{itemize}
\item FOR $i=1:N$
\item[-] Draw $\mathbf{x_k^i} \sim q(\mathbf{x_k}|\mathbf{x_{k-1}^i},\mathbf{z_k})$
\item[-] Evaluate the importance weights
$$\tilde{w}_k^i=w_{k-1}^i\frac{p(\mathbf{z_k|x_k^i})p(\mathbf{x_k^i|x_{k-1}^i})}{q(\mathbf{x_k^i|x_{0:k-1}^i,z_{1:k}})}$$
\item END FOR
\item FOR $i=1:N$
\item Normalizing weight:$w_k^i=\frac{\tilde{w}_k^i}{\sum_{j=1}^N \tilde{w}_k^j}$
\item END FOR
\end{itemize}
\end{algorithmic}
\end{algorithm}
\section{Resampling}
\textit{Degeneracy Problem}. A common problem associated with SIS particle filter is the degeneracy phenomenon where after a few recursive steps, all but one particle will have negligible weights. It implies that a large amount of computational effort is wasted in updating particles whose contribution is almost zero. A simple approach to resolve this problem is to increase $N$ but this will increase the computational cost which is unacceptable in practice. Instead, we introduce the concept \textit{effective sample size} $\hat{N_{eff}}$ which is evaluated as
\begin{equation}{~\label{Neff}}
\hat{N_{eff}}=\frac{1}{\sum_{i=1}^N(w_k^i)^2}
\end{equation}
Small $N_{eff}$ indicates severe degeneracy so the approach is to perform the \textit{resampling} when $\hat{N_{eff}}$ is below some threshold.  The idea of resampling is to eliminate the low-weighted particles and to concentrate on particles with large weights. It involves a mapping of random measure $\{x_{0:k}^i,w_k^i\}$ into a random measure $\{{x_{0:k}^i}^*,1/N\}$ with uniform weights and an efficient resampling algorithm named \textit{systematic resampling} is described in Algorithm~\ref{al:resampling}.~\cite{Ristic}
\begin{algorithm}
\caption{Resampling}
\label{al:resampling}
\begin{algorithmic}
\STATE $[{x_k^j,w_k^j}]=Resampling[{x_{k}^i,w_{k}^i}]$
\begin{itemize}
\item Find cumulative sum (CS) of the weights
\item Start from the bottom of CS: i=1
\item Draw a starting point $u_1 \sim \mathcal{U}[0,N_s^{-1}]$
\item FOR $j=1:N_s$
\item[-] Move along the CS: $u_j=u_1+N_s^{-1}(j-1)$
\item[-] WHILE $u_j>c_i$
\item[-] $i++$
\item[-] END WHILE
\item[-] Assign sample $x_k^j=u_j$
\item[-] Assign weight $w_k^j=N_s^{-1}$
\item END FOR
\end{itemize}
\end{algorithmic}
\end{algorithm}
\newline
Other possible resampling algorithms can be referred to ~\cite{hol2006resampling}. So far we have defined the main steps of a generic particle filter. The complete generic particle filter algorithm is summarized in Algorithm~\ref{al:General PF}.~\cite{Ristic}
\begin{algorithm}
\caption{SIS Particle Filter}
\label{al:General PF}
\begin{algorithmic}
\STATE $[{\mathbf{x_k^i},w_k^i}]=PF[{\mathbf{x_{k-1}}^i,w_{k-1}^i},\mathbf{z_k}]$
\begin{itemize}
\item Filtering via SIS ~\ref{al:SIS}
\item Calculate $\hat{N_{eff}}$ using ~\ref{Neff}
\item IF $\hat{N_{eff}}$ < $N_{Threshold}$
\item[-] Resampling
\item END IF
\end{itemize}
\end{algorithmic}
\end{algorithm}
And the simulation result using this algorithm is shown in the following figure
\begin{figure}[h!]{\label{fig:PF_systematicResampling}}
\centering
\includegraphics[width=1\textwidth]{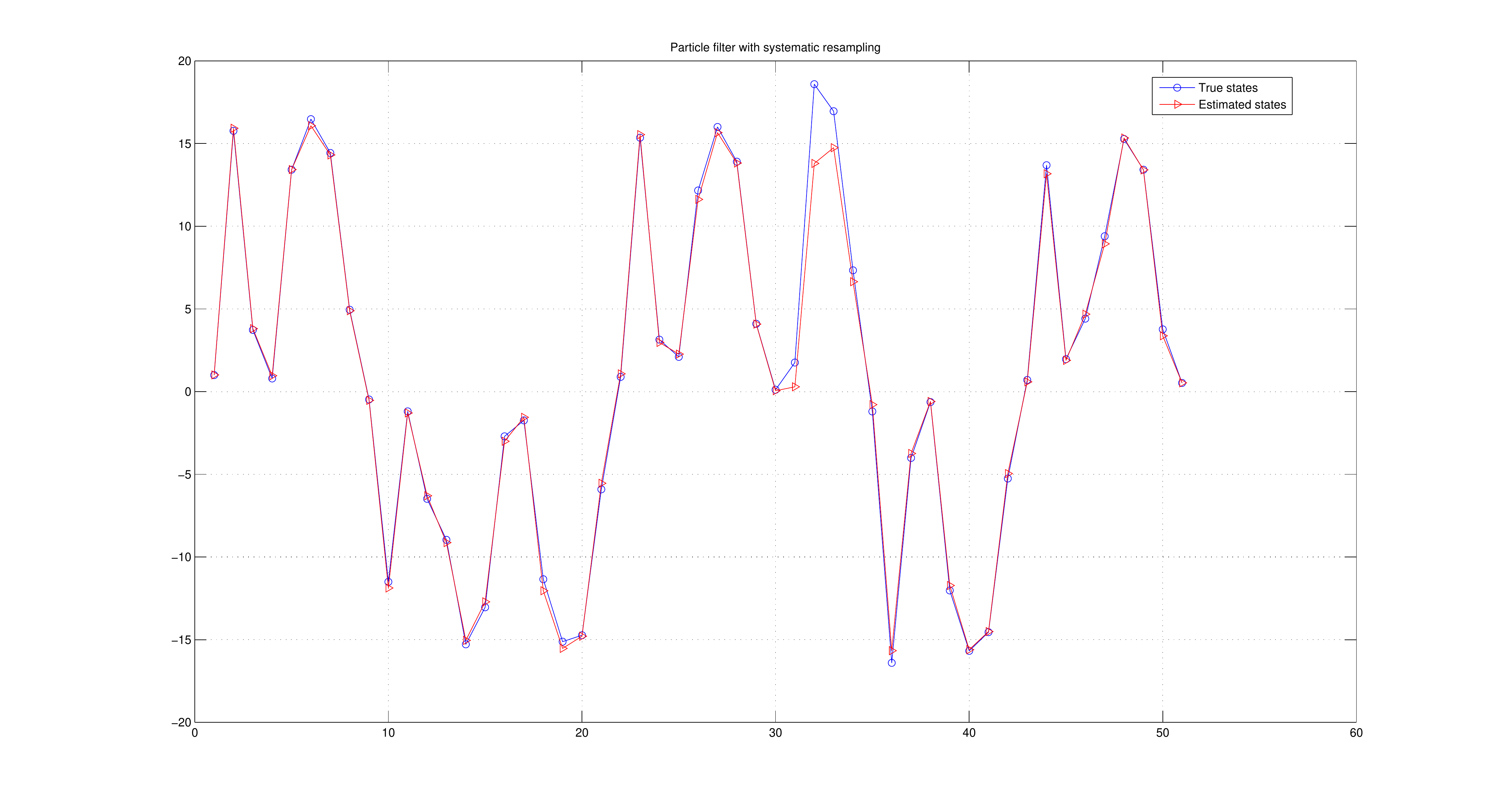}
\caption{Particle filter with systematic resampling}
\end{figure}
\section{Simulation}
Consider a non-linear SSM example using the Particle filter algorithm with systematic resampling as follows.
\begin{align*}
x_k &= \frac{x_{k-1}}{2}+\frac{25x_{k-1}}{1+x_{k-1}^2}+8cos(1.2(k-1))+w_{k-1}\\
z_k &= \frac{x_k^2}{20}+v_k
\end{align*}
where $w$ and $v$ are both Gaussian noise with zero mean and variance $var\_w, var\_v$ respectively. In the simulation, we use $500$ particles and track $50$ time steps.
\section{Summary}
In this chapter we described the generic sequential importance sampling algorithm which serves as a basis for most particle filters.
Compromising by the cost of high computational complexity, this generic particle filter can solve the non-linear state-space model with a good performance result.
However, there are some special cases of SIS algorithms which are derived by an appropriate choice of importance sampling density and/or modification of the resampling step.
Possible special particle filters are~\cite{arulampalam2002tutorial}
\begin{itemize}
\item sampling importance resampling (SIR) filter
\item auxiliary sampling importance resampling (ASIR) filter
\item regularized particle filter (RPF)
\end{itemize}
In accordance with the practical problem, we will select a suitable particle filter to be deployed.

\chapter{Gaussian Process}\label{ch:GP}
Kalman filter (Chapter~\ref{ch:KF}) and Particle filter (Chapter~\ref{ch:PF}) all rely on the condition that we have the deterministic state-space model structure. However in many cases, we deal with the problem that involves the SSM with uncertain structure. We are therefore required to jointly estimate the model structure as well as the state of the model. Rather than deciding that the unknown function relates to some specific models, a Gaussian process can represent the function flexibly, but rigorously, by letting the data decide the model structure. In this chapter, we will introduce how to use Gaussian processes for regression problems.
\section{Prediction Problem}
A typical prediction problem is that given some noisy observations of a dependent variable at certain values of the independent variable $x$, what the best estimate of the dependent variable at a new value $x_*$ is. This is modeled as
$$y_n = f_{w}(x_n) + \mathcal{N}(0,\sigma_n^2)$$

\section{Bayesian Inference}
The Bayesian approach is used for inference based upon the expression of knowledge in terms of probability distributions. Given the data and a specific model, we can deterministically make inferences using the rules of probability theory. Within the Bayesian approach to regression, we firstly infer the parameters $w$ of the model given the data and then to make predictions based on the chosen models and parameters.
\newline
We start by expressing prior beliefs about the model for the data in terms of a probability distribution over all possible function models, $p(M_{i})$. Then we express prior beliefs about the value of model parameters as $p(w|M_{i})$.
\newline
Including the data $x$ and $y$, we infer the parameters of the model given the data $$p(w|x,y,M_{i})=\frac{p(w|M_{i})p(y|x,w,M_{i})}{p(y|x,M_{i})}$$
where $p(w|x,y,M_{i})$ is the \textit{posterior}, $p(y|x,w,M_{i})$ is the \textit{likelihood}, $p(w|M_{i})$ is the \textit{prior} and $p(y|x,M_{i})$ is the \textit{evidence} or \textit{marginal likelihood}.
\newline
Next, we combine the evidence
$$p(y|x,M_{i})=\int p(w|M_{i})p(y|x,w,M_{i})dw$$
with prior belief and apply Bayes' theorem once more to find the model probability
$$p(M_{i}|x,y)=\frac{p(M_{i})p(y|x,M_{i})}{p(y|x)}$$
where $p(y|x)$ is the normalizing constant and this posterior distribution $p(M_{i}|x,y)$ allows us to rank different models.
\newline
Finally, we make the predictions of the future data relied on all of above equations.
$$p(y^*|x^*,x,y,M_{i})=\int p(y^*|w,x^*,M_{i})p(w|x,y,M_{i})dw$$
\newline
Despite Bayesian approach provides a uniquely optimal solution to the regression problem in theory, solutions may be difficult to find as in practice.
The fundamental difficulty of Bayesian approaches centers around the mathematical complexity where intractable integrals and awkward equations may often occur~\cite{gibbs1997bayesian}.

\section{Gaussian Processes}
Alternatively, Gaussian process techniques are introduced to formulate a Bayesion framework for regression~\cite{rasmussen2004gaussian} in a flexible and rigorous manner.
Initially we start with the basic multivariate Gaussian distribution (MVN)
$$p(x|\mu,\Sigma)=\textit{N}(\mu,\Sigma)=(2\pi)^{-D/2}|\Sigma|^{-1/2}\exp(-\frac{1}{2}(x-\mu)^T\Sigma^{-1}(x-\mu))$$
where the mean vector $\mu \in \mathbb{R}^D$ and the covariance matrix $\Sigma \in \mathbb{R}^{D\times D}$. As a generalization of of the MVN, a Gaussian process (GP) is
extending the $D$ dimensions into infinity which can be used to model a function which can be viewed as an aggregate for infinite quantity of random variables. The formal definition for a GP\cite{rasmussen2005gaussian} is as follows.
\newline
\newline
\textbf{Definition 1.}  \textit{A Gaussian process is a collection of random variables, any finite number of which have a joint Gaussian distribution.}
\newline
\newline
A Gaussian process (GP) is fully characterized by its mean function $m(x)$ and covariance function (also known as kernel) $k(x,x')$ which are defined as
\begin{align*}
m(x) &= \mathbb{E} [f(x)] \\
k(x,x') &= \mathbb{E} [(f(x)-m(x))(f(x')-m(x'))]
\end{align*}
and we write the Gaussian process as
$$f(x)\sim \mathcal{GP}(m(x),k(x,x'))$$

Note that the individual random variables in a vector from a Gaussian distribution are indexed by their positions in the vector instead of the time instants. For the Gaussian process it is the argument $x$ of the random function $f(x)$ that plays the role of index set: for every input x there is an associated random variable $f(x)$, which is the value of the stochastic function $f$ at that location.
\newline
Although it seems unwieldy to work with an infinitely long mean vector and an infinite covariance matrix, it turns out that the quantities that we are interested in computing require only working with finite dimensional objects. For any GP $f \sim \mathcal{GP}(m,k)$ we only put attention on a finite subset of function values $f=(f(x_{1}),f(x_{2}),\ldots,f(x_{n}))$ which follows a regular Gaussian distribution such that $$f \sim \mathcal{N}(\mu,\Sigma)$$ where $\mu=0, \Sigma _{ij}=k(x_{i},x_{j})$. To clarify the distinction between process and distribution we use $m$ and $k$ to the former and $\mu$ and $\Sigma$ for the latter. By using the properties of MVN we can make a prediction on $y^*$ based on the training pairs $\{\textbf{x},\textbf{y}\}$ and the test input $x^*$.

\section{Posterior Gaussian Process}
In the previous section we saw how to define distributions over functions using GPs. This GP will be used as a prior for Bayesian inference. We are usually not primarily interested in drawing random functions from the prior, but want to incorporate the knowledge that the training data provides about the function. Let us start with the simple special case where no noise is added on the observation. The joint distribution of the training outputs $f$ and the test outputs $f_*$ according to the prior is
\begin{equation*}
\begin{bmatrix}
f \\
f_*
\end{bmatrix}
\sim \mathcal{N}(
\begin{bmatrix}
m \\
m_*
\end{bmatrix}
,
\begin{bmatrix}
K(X,X) & K(X,X_*) \\
K(X_*,X) & K(X_*,X_*)
)
\end{bmatrix}
\end{equation*}
\newline
where we have $m$ for the training means and similarly $m_*$ for the test means. Also, we have $K$ for training set covariances, $K_*$ for training-test set covariance and $K_{**}$ for test set covariance.
\newline
\newline
\textbf{Lemma 1.} \textit{The formula for conditioning a joint Gaussian distribution is}~\cite{rasmussen2004gaussian}
\begin{equation*}
\begin{bmatrix}
x \\
y
\end{bmatrix}
\sim \mathcal{N}(
\begin{bmatrix}
a \\
b
\end{bmatrix}
,
\begin{bmatrix}
A & C \\
C^T & B

\end{bmatrix}
\Rightarrow
P(x|y) \sim \mathcal{N}(a+CB^{-1}(y-b),A-CB^{-1}C^T)
\end{equation*}
\newline
 Since we know the values for the training set $f$ we can obtain the conditional distribution of $f_*$ given $f$ as
\begin{align*}
f_*|f \sim \mathcal{N} & (m_*+K(X_*,X)K(X,X)^{-1}(f-m),\\
                       &  K(X_*,X_*)-K(X_*,X)K(X,X)^{-1}K(X,X_*))
\end{align*}
This is a prediction based on noise-free observations. In practice, it is more realistic modeling situations without the access to function values themselves. Instead, we only obtain the noisy observations thereof $y=f(x)+\mathcal{N}(0,\sigma_n^2)$.
\newline
Incorporating with the additive independent identically distributed (i.i.d.) Gaussian noise, we form a single kernel such that
$$cov(y)=K(X,X)+\sigma_n^2I$$
Thus we modify the joint distribution of the observed target values and the function values at the test locations under the prior as
\begin{equation*}
\begin{bmatrix}
y \\
y_*
\end{bmatrix}
\sim \mathcal{N}(
\begin{bmatrix}
m \\
m_*
\end{bmatrix}
,
\begin{bmatrix}
K(X,X)+\sigma_n^2I & K(X,X_*) \\
K(X_*,X) & K(X_*,X_*)
)
\end{bmatrix}
\end{equation*}
A Gaussian process posterior is
\begin{align*}
f(x^*)|x,y & \sim \mathcal{GP}(m_{post}(x),k_{post}(x,x')), where \\
m_{post}(x) & =m_*+ k(x^*,x)^T(K(x,x)+\sigma_n^2I)^{-1}y,\\
k_{post}(x,x') & = k(x^*,x^*)-k(x^*,x)^T(K(x,x)+\sigma_n^2I)^{-1}k(x^*,x)
\end{align*}
This leads us to the key predictive equations for Gaussian process regression
\begin{align}{\label{GP predictive distribution}}
y_*|x_*,x,y & \sim \mathcal{N}(m(y_*),cov(y_*)), where \\
     m(y_*) & = m_*+K(X_*,X)(K(X,X)+\sigma_n^2I)^{-1}(y-m),\\
   cov(y_*) & = K(X_*,X_*)+\sigma_n^2-K(X_*,X)(K(X,X)+\sigma_n^2I)^{-1}K(X,X_*)
\end{align}
Note that the variance is independent of the observed outputs $y$ and it is the difference between the prior variance and a positive term, representing the information the observation gives us about the function.
\newline
Consider an example of the Gaussian process.
\newline
\textit{Example.}
\begin{align*}
y &=f(x)+\mathcal{N}(0,\sigma_n^2)\\
f &\sim \mathcal{GP}(0,k(x,x'))\\
k(x,x') &= \exp(-\frac{1}{2}(x-x')^2)
\end{align*}
\textit{Solution.}
Zero mean Gaussian process prior leads to the Gaussian predictive distribution:
\begin{align}{\label{GP zero mean prediction}}
y^*|x^*,x,y \sim \mathcal{N} & (k(x^*,x)^T(K(x,x)+\sigma_n^2I)^{-1}y,\\
                       &  k(X^*,X^*)+\sigma_n^2-k(x*,x)^T(K(x,x)+\sigma_n^2I)^{-1}k(x^*,x))
\end{align}
A practical implementation of Gaussian process regression is shown in the figure and the MatLab code is in Appendix
\begin{figure}[h!]{\label{fig:GPR}}
\centering
\includegraphics[width=1\textwidth]{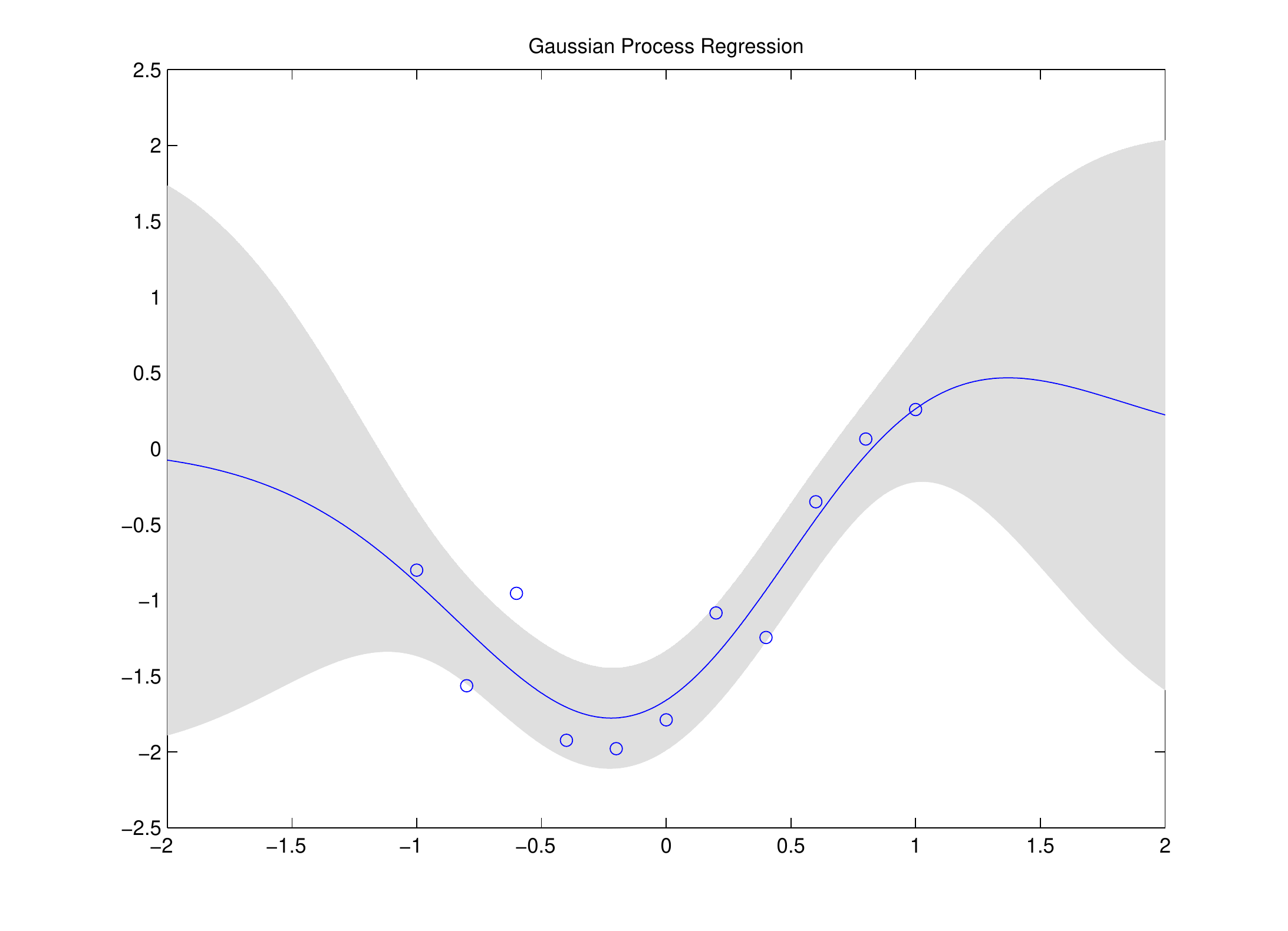}
\caption{Gaussian process regression}
\end{figure}

Instead of directly inverting the matrix, Cholesky decomposition of a matrix can be used since it is faster and numerically more stable. A good feature of GP is that it gives both the predictive mean (the blue curve) and $95\%$ posterior confidence region (the grey shaded area).
\newline

Note that in the result (~\ref{GP zero mean prediction}), the mean prediction is a linear combination of observations $y$ when the prior mean is zero. This is often referred to as a linear predictor~\cite{rasmussen2005gaussian} and this mean equation can be re-written as $$\mu(y^*)=\sum_{i=1}^n \alpha_{i}k(x_i,x^*)$$
where $\alpha_{i}=(K+\sigma_n^2I)^{-1}y$ and $K$ is the compact form of $K(x,x')$. This can be seen as a linear combination of $n$ kernel functions, each one centered on a training point. Intuitively, although the GP defines a joint Gaussian distribution over all of the $y$ variables, one for each point in the index set $\mathcal{X}$, for making prediction at $x_*$ we only care about the $(n+1)$ dimensional distribution defined by the $n$ training points and the test point.~\cite{rasmussen2005gaussian} This prediction can be given by conditioning this $(n+1)$ dimensional distribution on the observations as shown above.
\newline

\section{Training a Gaussian Process}
Now a question left is which kernel function to choose and how to determine the hyperparameters. In the light of training data, we need to find reliable prior information about the training data set with prior mean and covariance functions specified before making regression. However, the availability of such detailed prior information is not valid normally. Referred as the training of GP, we need to form a mean and kernel function as the GP prior and in the light of observations, we calculate the appropriate hyperparameters within the function.
\newline
\textit{Task 1. Form Covariance Function}
\newline
There are a set of well known covariance functions which are appropriate in different cases.\cite{rasmussen2005gaussian}
\begin{itemize}
\item Long-term smooth trend - Square Exponential
$$k_1(x,x')=\theta_{1}^2 \exp(-(x-x')^2/\theta_{2}^2)$$
\item Seasonal trend - Quasi-periodic Smooth
$$k_2(x,x')=\theta_{3}^2 \exp(-2\sin^2(\pi(x-x'))/\theta_{5}^2)\times \exp(-\frac{1}{2}(x-x')^2/\theta_{4}^2)$$
\item Short- and medium-term anomaly - Rational Quadratic
$$k_3(x,x')=\theta_{6}^2(1+\frac{(x-x')^2}{2\theta_{8}\theta_{7}^2})^{-\theta_{8}}$$
\item Noise - Independent Gaussian and Dependent
$$k_4(x,x')=\theta_{9}^2 \exp(-\frac{(x-x')^2}{2\theta_{10}^2})+\theta_{11}^2\delta_{xx'}$$
\end{itemize}
By linearly combining them we obtain a comprehensive covariance function that utilizes the comparative advantages and compensates the drawbacks to large extent.
$$k(x,x')=k_1(x,x')+k_2(x,x')+k_3(x,x')+k_4(x,x')$$
\textit{Task 2. Find Hyperparameters}
\newline
For a Gaussian Process, $$f \sim \mathcal{GP}(m,k)$$
the mean and covariance functions are parameterized in terms of hyperparameters $\theta = \{\theta_m, \theta_k\}$ where $\theta_m$ and $\theta_k$ indicate hyperparameters of mean and covariance functions respectively.
In order to find the values for these hyperparameters, we compute the probability of the data given the hyperparameters by introducing the log marginal likelihood (or evidence) since by assumption the distribution of the data is Gaussian:
$$L=logP(y|x,\theta)=-\frac{1}{2}(y-m)^TK^{-1}(y-m)-\frac{1}{2}log|K|-\frac{n}{2}log(2\pi)$$
\newline
Then we can find the values of hyperparameters which optimizes the marginal likelihood based on its partial derivatives:
\begin{align*}
\frac{\partial L}{\partial \theta_m} &=-(y-m)^TK^{-1}\frac{\partial m}{\partial \theta_m}\\
\frac{\partial L}{\partial \theta_k} &=\frac{1}{2}(y-m)^TK^{-1}\frac{\partial K}{\partial \theta_k}K^{-1}(y-m)-\frac{1}{2}trace(K^{-1}\frac{\partial K}{\partial \theta_k})
\end{align*}
The log marginal likelihood form consists of three terms: The first term $-\frac{1}{2}(y-m)^TK^{-1}(y-m)$ is a negative quadratic and plays the role of a data fit measure as it is the only term which depends on the training set output values $y$. The second term $-\frac{1}{2}log|K|$ is a complexity penalty term, which measures and penalizes the complexity of the model. The third term is a log normalization term that is independent of the data. Note that the tradeoff between penalty and data fit - Occam's Razor - in the GP model is automatic.~\cite{rasmussen2004gaussian} There is no weighting parameter which needs to be set by external method and this feature has great practical importance since it simplifies training.
\section{Summary}
In this chapter we have introduced the basic concept of Gaussian process with its application on how to solve the regression problem with a GP flexibly as well as rigorously. Moreover, we illustrated multiple common-used kernel functions and the method deployed to resolve the hyperparameters associated.

\chapter{Conclusion and Future Work}\label{ch:conclusion}
Following the methodology-oriented research principle, fundamental knowledge of classical approaches to solve the state-space model with known structure are learnt. In thesis B, Gaussian process prior is to be incorporated with particle filter to solve some practical problem in wireless communications like channel estimation, which involves a non-linear state-space model with structure uncertainty. In analogy, after drawing a series of discrete points in the paper, we will find a proper line to connect those points to contribute to an agreeable outcome.
\newline
Future work may include combining Gaussian process prior within state-space model to solve some practical problems in wireless communications. One possible problem is the channel tracking in relay networks~\cite{nevat2010channel} where the system model is illustrated in the following figure.
\begin{figure}[h!]{\label{fig:Relay network}}
\centering
\includegraphics[width=1\textwidth]{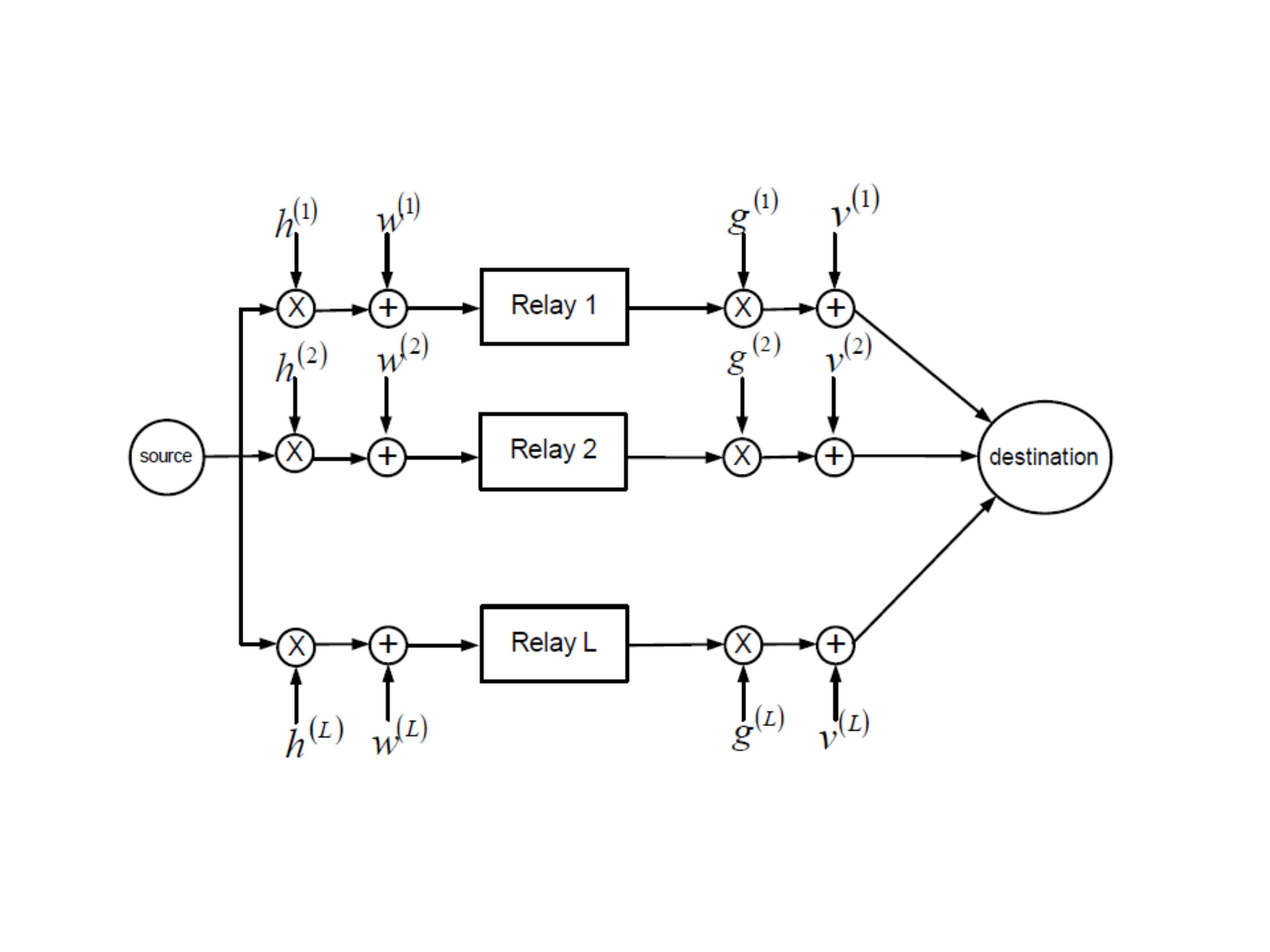}
\caption{Relay network system model}
\end{figure}
If we assume the relay function is unknown, then this channel tracking problem involves a non-linear state-space model with parameters estimation. In this case, we need to incorporate Gaussian process for the function estimation with particle filter for the channel state information recovery.

\addcontentsline{toc}{chapter}{Bibliography}
\bibliographystyle{plain}
\bibliography{pubs}
\appendix
\addcontentsline{toc}{chapter}{Appendix}

\chapter*{Appendix 1 - Kalman Filter for AR2 model}
\begin{verbatim}
clear all;
close all;
clc
% System Model
% [x(n+1);x(n)] = [cos(2*pi*f) -1; 1 0]*[x(n);x(n-1)];
% y(n) = [1 0]*[x(n);x(n-1)] + v(n);

f = 0.05;
theta=1;
F = [2*cos(2*pi*f) -1;1 0];
H = [1 0];
R = 0.1;      % Measurement noise covariance
Q = 0.1;      % Process noise covariance
N = 300;

x_state=zeros(2,N);     % the real state
x_hat=zeros(2,N);       % estimate state
P=zeros(2,2,N);         % covariance error matrix N*(2,2)
x_priori=zeros(2,N);    % aprior estimate state
K=zeros(2,1,N);         % Kalman gain

% System model setup
x_state(:,1)=[sin(theta);0];
x_state(:,2)=[sin(2*pi*f + theta);sin(theta)];
for t=3:N
    x_state(:,t)=F*x_state(:,t-1);
end
for t=1:N
    v=normrnd(0,sqrt(R),1,1);
    y(t) = H*x_state(:,t) + v;
end

% Initial guess
x_hat_initial=[sin(theta);0];   % random initial state estimate
P_initial = [1 0; 0 1];
% First round of Kalman Filter
[x_hat(:,1), x_prior(:,1), P(:,:,1), K(:,:,1)] = KalmanFilter(x_hat_initial, P_initial, y(1), F, H, Q, R);

for t=2:N
    [x_hat(:,t), x_prior(:,t), P(:,:,t), K(:,:,t)] = KalmanFilter(x_hat(:,t-1), P(:,:,t-1), y(t), F, H, Q, R);
end
t=1:N;
figure
plot(t,x_state,'b',t,y,'k.',t,x_hat,'r')
grid on
\end{verbatim}
\chapter*{Appendix 2 - Kalman Filter Function}
\begin{verbatim}
function [x, x_prior, P, K ] = KalmanFilter(x, P, z, F, H, Q, R)
 % Projection
x_prior = F*x;
P = F * P * F' + Q;

% Kalman gain
K = P*H'*inv(H*P*H'+R);
% Update estimate
x = x_prior + K*(z-H*x_prior);

% Update covariance
dimension=size(K*H,1);
P=(eye(dimension)-K*H)*P;
end
\end{verbatim}
\chapter*{Appendix 3 - Particle Filter}
\begin{verbatim}
%% Clean up
clear all
close all
clc

%% Set up problem parameters
randn('state',1) % initialize Gaussian random number generator
rand('twister',1) % initialize uniform random number generator
N = 500; % # of particles
K = 50; % # of timesteps
T = 0:K; % time vector

%% Generate data

vr_w = 0.1; % variance of Gaussian noise parameter w
vr_v = 0.5; % variance of Gaussian noise parameter v
x0 = 0.1; % initial state value
P0 = 0.1; % initial state variance
x = x0;

% generate state and measurement vectors
for i=2:K+1
    x(i) = x(i-1)/2 + 25*x(i-1)/(1+x(i-1)^2) + 8*cos(1.2*(i-1)) + sqrt(vr_w)*randn;
end
z = (x.^2)./20 + sqrt(vr_v).*randn(size(x));

%% Initialize particle filter
% The set of particles and their weights are denoted by j Xk j and j Wk j
% respectively, while j mn j is the mean of the particle distribution. It is
% assumed that j x0 j is known and we chose out initial state pdf to be a
% Gaussian distribution about j x0 j with the variance P0.

Xk = x0 + randn(1,N)*sqrt(P0); % initial particle population
Wk = (1/sqrt(2*pi*P0))*exp(-(Xk-x0).^2/(2*P0)); % initial weight dist
Wk = Wk/sum(Wk); % weight normalization
mn = Xk*Wk'; % initial particle mean
maxX = max(Xk);
minX = min(Xk);

%% Run particle filter

for t=2:K+1
    %Propagate particles
    Xk = Xk./2 + 25*Xk./(1+Xk.^2) + 8*cos(1.2*(t-1)) + sqrt(vr_w)*randn;

    %Update weights
    %posterior pdf
    Wk = Wk.*((1/sqrt(2*pi*vr_v))*exp(-(z(t)-(Xk.^2)./20).^2/(2*vr_v)));
    Wk = Wk/sum(Wk);

    %Infer particle mean (aggregate state estimate)
    maxX(t) = max(Xk);
    minX(t) = min(Xk);
    mn(t) = Xk*Wk';

    %Multinomial resampling
    n_thr = 0.25*N;
    n_eff = 1/(sum(Wk.^2));
    if n_eff<n_thr
        cs = cumsum(Wk); % generate cumulative sum
        % vector for the weights (CSW)
        for i=1:N
            indx = min(find(cs > rand)); % find CSW index for which the
            % CSW just exceeds the random number
            Xk(i) = Xk(indx); % replicate the corresponding
            % particle in the new population
        end
        Wk = ones(size(Wk))/N; % assign uniform weights to
        % resampled particles

    end
end
plot(x)
hold on
plot(mn,'g')
\end{verbatim}

\chapter*{Appendix 4 - Gaussian Process Regression}\label{Appendix:GPR}
\begin{verbatim}
% Posterior prediction
%%
clear all;
close all;
clc
%% Training data
var_n=0.1;  % noise variance
var=1;      % kernel hyperparameter
l=0.5;      % kernel hyperparameter
training_x=[-1:0.2:1];
number_data=length(training_x);     % number of training data
K=se_cov(training_x,training_x,var,l);  % covariance matrix
mean_y=zeros(number_data,1);
training_y=mvnrnd(mean_y,K);        % y~N(0,K)
training_y=training_y'+sqrt(var_n)*randn(number_data,1); % y=f(x)+noise
%% Predict test data
test_x=[-2:0.001:2];    % test input
mean_test_y=zeros(1,length(test_x));
for i=1:length(test_x)
    mean_test_y(i)=se_cov(test_x(i),training_x,var,l)'*inv(K+var_n*eye(number_data))*training_y;    % mean of test output
    var_test_y(i)=se_cov(test_x(i),test_x(i),var,l)-se_cov(test_x(i),training_x,var,l)'*inv(K+var_n*eye(number_data))*se_cov(test_x(i),training_x,var,l);   % variance of test output
end
%% Plot
plot(test_x,mean_test_y,'r',training_x,training_y,'ob')
cf_upper=mean_test_y+2*sqrt(var_test_y);
cf_lower=mean_test_y-2*sqrt(var_test_y);
f = [cf_upper; flipdim(cf_lower,1)];
fill([test_x; flipdim(test_x,1)], f, [7 7 7]/8, 'EdgeColor', [7 7 7]/8)
hold on
plot(test_x,mean_test_y,training_x,training_y,'ob')
\end{verbatim}
\chapter*{Appendix 5 - Covariance Function}\label{Appendix:kernel}
\begin{verbatim}
% Calculate covariance funciton
function K = se_cov(x, y,var,l);
K=zeros(length(x),length(y));
for i=1:length(x)
    for j=1:length(y)
        K(i,j)=var*exp(-0.5/l*(x(i)-y(j))^2);
    end
end
K=K';
end
\end{verbatim}
\end{document}